\documentstyle[12pt,psfig]{article}
\begin{document}
\title{EROS/MACHO Gravitational Microlensing Events Toward LMC in Evans
Halo Model}
\author{Sohrab Rahvar\footnote{E-mail:rahvar@mehr.sharif.edu}\\
Department of Physics, Sharif University of Technology,\\
 P.O.Box 11365--9161, Tehran, Iran\\
Institute for Studies in Theoretical Physics and Mathematics,\\
P.O.Box 19395--5531, Tehran, Iran}
\maketitle
 \abstract
 { After a decade of gravitational microlensing experiments, 13 to 17
 events by MACHO (depending on quality) and two events by
EROS have been detected. All of those have been observed in the
direction of Large Magellanic Cloud. We use Evans spherically
symmetric halo model to study the rate of microlensing events.
The expected number of events in this model obtain by using EROS
and MACHO observational efficiencies. We compare our numbers with
the observed events to obtain the fraction of halo that is made
by compact objects. It is shown that results derived from two
experiments are in good agreement with each other and MACHOs,
comprise only a fraction (depending on the model) of Milky Way
halo. The results are also compatible with the White Dwarfs
population studies in the Hubble Deep Field. \maketitle
\section{Introduction}
The existence of dark matter becomes apparent by studying the
rotational curve of spiral galaxies \cite{fab79} \& \cite{tri87}.
Recent results of optical and 21 cm band observations shows that
for thousands of spiral galaxies, the Keplerian rotational
velocity beyond the luminous radius remains constant \cite{per96}.
Diffusion emission in X-rays from elliptical galaxies and dynamics
of cluster of galaxies also shows that there should exist a halo
structure around spiral galaxies. Various dark matter candidates
such as, baryonic dark matter and exotic dark candidates such as
Axions, massive neutrinos, WIMPs and Super-symmetric particles
have been proposed. However, simulations show that there is a
discrepancy between expected rotational velocity due to cold
dark matter of halo and observed velocity curves \cite{moo94} \& \cite{nav96}.\\
In the case of baryonic dark matter, Big Bang nucleosynthesis
models give $\Omega_{B} h^2 = 0.02$ \cite{cop95} \& \cite{bur98}
while measuring the mass of luminous stars obtain $\Omega_{lum}$
of universe to be around $0.004$ \cite{fuk95}. Comparing
$\Omega_{B}$ with $\Omega_{lum}$ ($\Omega_B\gg\Omega_{\it lum}$),
provides an other evidence for the existence of baryonic dark
matter. One of the types of baryonic dark matter could be in the
form of Massive Astrophysical Compact Halo Objects (MACHO). These
objects due to their light masses are obscure. Neutron stars and
black holes as dark objects can also be considered in this
category. The pioneer work of using gravitational microlensing
technique for detection of compact objects was proposed by
Paczy\'nski \cite{pac86}. Since his proposal, microlensing
searches have turned very quickly into reality and some groups
like AGAPE, DUO, EROS, MACHO, OGLE and PLANET have contributed to
this field . After one decade, the final results of experiments
compared with the theoretically expected results, provides a
constraint on the fraction of halo in the form of compact objects.
This results strongly depends on the galactic model and mass
function of MACHOs. Using the standard model for Milky Way and
delta mass function for MACHOs, EROS and MACHO groups could
obtain the fraction of halo that is made of compact objects. EROS
result, puts a strong constraint on the fraction of halo made of
objects in the range $\left[ 10^{-7} M_{\odot}, 4
M_{\odot}\right]$, excluding at $ 95\%$ C.L. that more than $40
\%$ of the standard halo is made of objects with up to one solar
mass \cite{spi01}. The analysis of 5.5 years observation of LMC by
the MACHO group estimates that the halo mass fraction in the form
of compact halo objects is about $20\%$  \cite{alc00}.\\
Here, we use the most general spherically symmetric model for the
halo of spiral galaxies that interprets Keplerian rotational
curves\cite{eva94}. The expected distribution of gravitational
microlensing events obtain by applying the efficiency of EROS and
MACHO experiments in these models. Here we use power law mass
function for the compact halo objects and mass function of disk
derived from HST observation. We compare the expected rate of
events with those observed experimentally to evaluate the
fraction of halo made by MACHOs.\\
The organization of the paper is as follows. In Sect. 2, we
introduce the basics of gravitational microlensing and obtain
relation between the optical depth and the rate of microlensing
events. In Sect. 3 we introduce Evans model for galactic
structure and in Sect. 4 we estimate by Monte-Carlo simulation
the expected rate of microlensing events toward LMC in six
different galactic models. We then calculate the fraction of halo
made by MACHO in different models. The results are discussed in
Sect. 5.
\section{Basics of gravitational microlensing}
In this section we present the main feature of gravitational
microlensing and in particular, we obtain relation between optical
depth and the rate of events. For review, see ( \cite{pac96},
\cite{rou97}, \cite{gou00}, \& \cite{jet97}).\\
According to general relativistic results, a given light ray
bends near a massive star. Considerable gravitational lensing
occurs when the line of sight between us and a background star
passes near enough a lens. Since the deflection angle in the case
of microlensing is too small (taking into account the resolution
of present apparatus), it is impossible to distinguish two images
that are produced due to the gravitational lensing, thus the
effect is only on the brightness magnification of the background
star. This magnification is given by
\begin{equation}
A(t) = \frac{{u(t)}^2 + 2}{u(t)\sqrt{{u(t)}^2 + 4}},
\end{equation}
where ${u(t)}^2 = {u_0}^2 + \left( \frac{t - t_0}{t_E}\right)^2$
is the impact parameter, normalized by Einstein Radius
\begin{equation}
{R_E}^2 = \frac{4GMD_{os}}{c^2}x(1-x).
\label{einr}
\end{equation}
In the definition of Einstein radius, $D_{ol}$ and $D_{os}$ are
the distance of the lens and source from the observer and $x$ is
the fraction of these two terms, ($x = \frac{D_{ol}}{D_{os}}$).
Definition of an event is given by constraint on maximum
magnification with $A_{max}>1.34$ or in
other word $u<R_E$.\\
One of the crucial relation in statistical analysis of
microlensing experiments is between the optical depth and the
rate of events. Taking a snapshot from the background stars, the
probability of existence the projected background stars inside
Einstein Radius of lenses in the lens plane along our line of
sight is the definition of optical depth. The optical depth of
Microlensing event in the range of $[M,M+dM]$ can be obtain by:
\begin{eqnarray}
d\tau(x;M,M+dM) &=& \frac{\pi
R_E^2}{A}\left(\frac{\rho(x;M,M+dM)}{M}\right)\times\left(A\cdot
dx\right). \\
\label{dtau} d\tau(x;M,M+dM) &=& \frac{4 \pi
GM}{c^2}D_{os}^2x(1-x)n(x)g(M) dM dx,
\end{eqnarray}
where, $dn = \frac{\rho(x;M,M+dM)}{M}$ is the number density of
lenses in the range of $[M,M+dM]$ and can be written in the terms
of mass function $g(M)$ and total density of compact object
$n_{total}$ as follows: $dn(M,M+dM) =n_{total}g(M)dM$,where the
mass function is normalized to one. It is seen in Eq.(\ref{dtau})
that optical depth is independent of mass function of MACHOs and
it is only function of the density distribution of matter. The
rate of events per year per the number of background stars
$\Gamma$, as an observable quantity depends on the optical depth.
The expected number of events in the range of $[M,M+dM]$ obtain:
\begin{equation}
\label{dn} dN_{exp}(x;M,M+dM) = \left( \frac{2 R_E(x)
\left<v_t(x)\right> T_{obs}}{A(x)} \right)\times \left( n(x) g(M)
dM \times A \cdot dx \right) \times N_{bg},
\end{equation}
where, $T_{obs}$ and $N_{bg}$ are the monitoring time and the
number of background stars, $T_{obs}\times N_{bg}$ is called the
exposure time and $\left<v_t(x)\right>$ is the mean transverse
velocity of lenses with respect to the line of sight. The mean
transverse velocity obtain by velocity distribution function,
$f_v(x)$ as follows:
\begin{equation}
\label{vt} \left<v_t\right> = \int v_t(x)f_v(x)d^3v,
\end{equation}
where, the distribution function of velocity is normalized to one.
The first term on the right hand side of Eq.(\ref{dn}) represents
the ratio of spanned tube by a lens to the projection of
background stars zone on the deflector plane, call it $A(x)$. The
Second term denotes the number of lenses inside $x$ and $x+dx$
and the last term shows the number of background stars. By
definition of $d\Gamma = \frac{d N_{exp}(x)}{N_{bg} T_{obs}}$ and
using Eq.(\ref{vt}), the rate of events is given by:
\begin{equation}
\label{dgam} d\Gamma(x;M,M+dM) = 2 (n(x) g(M) dM)
R_E(x)^2dx\int\frac{v_t(x)\epsilon(t_E)}{R_E(x)}f_v(x)d^3v,
\end{equation}
where, $\epsilon(t_E)$ is the efficiency of observation. By
substituting Eq.(\ref{dtau}) in Eq.(\ref{dgam}) and using the
definition of Einstein crossing time $t_E = \frac{R_E}{v_t}$,
relation between the rate of microlensing and the optical depth
obtain as follows:
\begin{equation}
\label{dgam2} d \Gamma(x) = \frac{2}{\pi} d\tau \int
\frac{f(x,v)\epsilon(t_E)}{t_E(x,v)}d^3v.
\end{equation}
In the right hand side of Eq.(\ref{dgam2}), the value of integral
is equal to the mean value of $\epsilon(t_E)/t_E$ ( $f_x(v)$ is
normalized to one). Rewriting Eq.(\ref{dgam2}) yields:
\begin{equation}
\label{gam3} d\Gamma(x) = \frac{2}{\pi} d \tau(x)
\left<\frac{\epsilon(t_E)}{t_E(x)}\right>.
\end{equation}
We take integral with respect to $x$, relation between rate of
events and optical depth is given by:
\begin{equation}
\label{gam4} \Gamma = \frac{2}{\pi}\tau \left<
\frac{\epsilon}{t_E} \right>.
\end{equation}
Here, we are interested in to know the rate of events for a given
optical depth. The theoretical rate of events
$(\frac{d\Gamma}{dt_E})$ is available by a Monte-Carlo simulation
based on the geometrical distribution of matter, velocity
distribution and the mass function of lenses. The observational
efficiency is also given as a function of event duration,
depending on experimental setup and the strategy of observation.
One can multiply these two distributions to estimate the expected
number of microlensing events as follows:
\begin{equation}
\Gamma = \int\frac{d\Gamma}{dt_E}\epsilon(t_E)dt_E
\end{equation}
In the next section we introduce different galactic models and
obtain the expected rate and the total number of microlensing
events toward Large Magellanic Cloud.

\section{Galactic models}
As mentioned in the last section, to calculate the rate of
events, we need to know the density and velocity distributions
and the mass function of lenses in the galactic model. The
galactic structure falls into three different parts of the bulge,
disk and halo. Our aim is to obtain the contribution of each part
in the optical depth and the rate of events toward our line of
sight (Large Magellanic Cloud). Since the contribution of the
bulge on optical depth toward this direction is negligible, we
ignore its contribution in our calculation. In what follows we
consider the structure of disk and halo.
\subsection{ Galactic disk }
The density profile of galactic disk can be given by a double
exponential in cylindrical galactic coordinate system $(R,z)$ as
follows \cite{bin87}:
\begin{equation}
\rho(R,z) = \frac{\Sigma_{\odot}}{2h} \exp\left[ -\frac{R -
R_{\odot}}{R_d} \right] \exp\left[-\frac{|z|}{h}\right]
\end{equation}
where $R_d \sim 3.5 kpc$ is in the order of disk radius and $h$
and $\Sigma_{\odot}$ represent the thickness and column density of
disk respectively. In our analysis we consider two types of disk
which are called thin and thick disk models. The thin disk mainly
is made up by the population of stars and gas. The column density
and thickness of disk in this model near the sun is about
$\Sigma_{\odot} \sim 50 M_{\odot} pc^{-2}$ and $h=0.32 kpc$. In
the case of thick disk, dark matter contribution also is taken
into account. Here, we choose $\Sigma_{\odot} \sim 100 M_{\odot}
pc^{-2}$ and $h = 1kpc$. The rotational velocity of disk also is
given by \cite{bin87}:
\begin{equation}
{V_{disk}(R)}^2 = 4 \pi G \Sigma_{\odot} R_d e^{R_{\odot}/R_d}y^2
[I_0(y)K_0(y) - I_1(y)K_1(y)],
\end{equation}
where $ y = R/{(2R_d)}$ and $I_n$ and $K_n$ are the modified
Bessel functions.  The velocity distribution for two models of
disk is shown in Figure(\ref{vel2}).
\begin{figure}[htbp] 
\vspace*{8pt} \centerline{\psfig{file=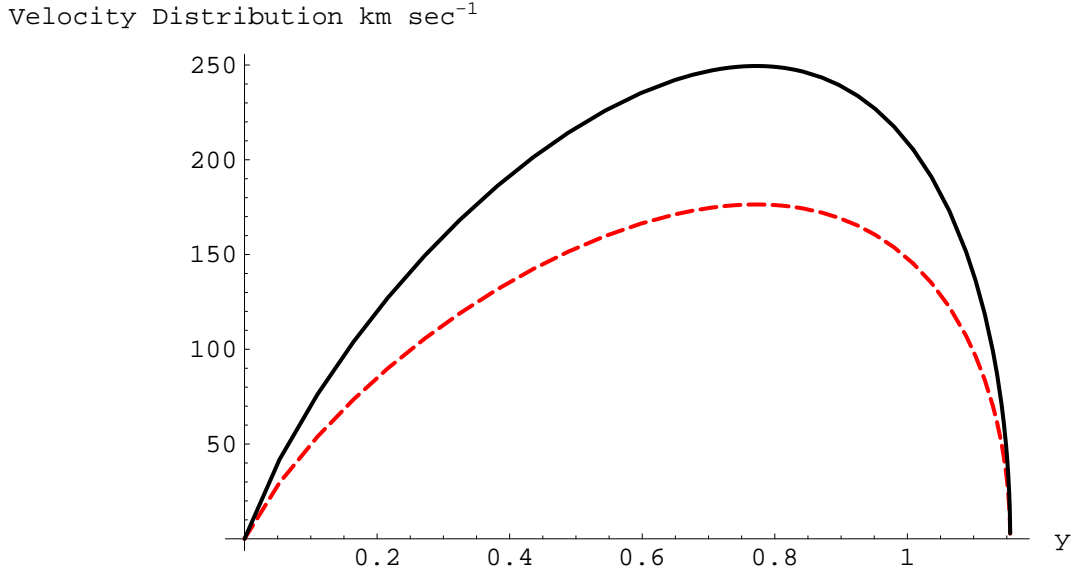}} \vspace*{8pt}
\caption{ Thick line represents the rotational velocity of thick
disk and the dashed line shows this distribution for thin disk,
x-axis indicates the distance of disk from the center of galaxy
in terms of $y =2R/R_d$.} \label{vel2}
\end{figure}
\subsection{ Power law  halo model }
The largest known set of axisymmetric model is called the "power
law galaxies". The density of halo in the cylindrical coordinate
system is given by:
\begin{equation}
\rho(R,z) =\frac{{V_a}^2{R_c}^{\beta}}{4\pi G q^2}\times
\frac{{R_c}^2(1+2q^2) + R^2(1-\beta q^2) +
z^2[2-(1+\beta)/q^2]}{({R_c}^2 + R^2 + z^2/q^2)^{(\beta+4)/2}},
\label{rho}
\end{equation}
where $R_c$ is the core radius and $q$ is the flattening
parameter, which is the axis ratio of concentric equipotential.
$q=1$ represents a spherical $(E0)$ halo and $q \sim 0.7$ gives
an ellipticity of about $E6$. The parameter of $\beta$ determines
whether the rotational curve asymptotically rises, falls or is
flat. At large distance $R$ in the equatorial plane, the
rotational velocity is given by $V_{circ}\sim R^{-\beta}$.
Therefore $\beta = 0$ corresponds to flat the rotational curve,
$\beta<0$ is a rising rotational curve, and $\beta>0$ is falling.
The parameter $V_a$ determines the overall depth of the potential
well and hence gives the typical velocities of lenses in the
halo. The density distribution in the spherical symmetric models
$(q=1)$ for an observer that located at the position of sun and
observes in the line of sight of Large Magellanic Cloud ({\it
l}=280, {\it b}=-33) obtain by:
\begin{equation}
\rho(r) = \frac{V_a^2 R_c^\beta}{4\pi G}\times
\frac{3R_c^2+(1-\beta)[R_0^2+r^2-2rR_0\cos(l)\cos(b)]}
{[R_c^2+(1-\beta)(R_0^2+r^2-2rR_0\cos(l)\cos(b))]^{(\beta+4)/2}}.
\end{equation}
The velocity dispersion of lenses relate to the center of galaxy
in the cylindrical coordinate system is given as follows
\cite{eva94}:
\begin{eqnarray}
{\sigma_r}^2 &=& {\sigma_z}^2 = \frac{{V_a}^2{R_c}^{\beta}}{2(1 + \beta)}\frac{1}{({R_c}^2 + R^2+z^2/q^2)^{\beta/2}} \nonumber\\
&\times& \frac{2q^2 {R_c}^{\beta} + (
1-\beta)q^2R^2+z^2[2-(1+\beta)/q^2]}{{R_c}^2(1+2q^2) + R^2(1-\beta
q^2)
+ z^2[2-(1+\beta)/q^2]},\\
 {\sigma_{\phi}}^2 &=& \frac{{V_a}^2{R_c}^{\beta}}{2(1 + \beta)}\frac{1}{({R_c}^2 + R^2+z^2/q^2)^{\beta/2}}
 \\ \nonumber
&\times& \frac{2q^2 {R_c}^{\beta} +
[2+2\beta-(1+3\beta)q^2]R^2+z^2[2-(1+\beta)/q^2]}{{R_c}^2(1+2q^2)
+ R^2(1-\beta q^2) + z^2[2-(1+\beta)/q^2]}.
\end{eqnarray}
Like the distribution of matter, we want to obtain the velocity
distribution related to the new frame at the position of sun. The
distribution of transverse velocity in the lens plane which we
denoted it by $v_t$, is given by taking integral along the line
of sight. For the spherical symmetric models $(q=1)$ the
transverse velocity distribution in the lens plane is
\begin{equation}
f(v_t)v_tdv_t =
\frac{1}{\sigma^2}\exp(-\frac{v_t^2}{2\sigma^2})v_tdv_t,
\end{equation}
where
\begin{eqnarray}
\sigma^2 &=& \frac{{V_a}^2{R_c}^{\beta}}{2(1 +
\beta)}\frac{1}{[{R_c}^2 + (1-\beta)(R_0^2+r^2-2rR_0\cos({\it l})
\cos({ \it b}))]^{\beta/2}} \nonumber\\
&\times& \frac{2{R_c}^{\beta} +
 (1-\beta)(R_0^2+r^2-2rR_0\cos({\it l})
\cos({ \it b}))}{3{R_c}^2 +(1-\beta)(R_0^2+r^2-2rR_0\cos({\it l})
\cos({ \it b}))}.
\end{eqnarray}
Here in our study, we take into account six galactic model
(\cite{alc94}, \cite{ren96} \& \cite{pal97}) with the following parameters:\\
Model 1a: standard halo and thin disk. \\
Model 1b: standard halo and thick disk.\\
Model 2a: power law model (q=1 and $\beta=0$) and thin disk.\\
Model 2b: power law model (q=1 and $\beta=0$) and thick disk. \\
Model 4: spherical halo with asymptotic decreasing  rotational
velocity (q=1 and $\beta=0.2$) and thin disk.\\
Model 6: spherical halo with flat rotational curve ( q =1 and
$\beta = 0$) and intermediate disk. \\
Table (1) shows the parameters of these models. \\
\begin{table*}
\begin{center}
\begin{tabular}{lclclclclclclclclc|}
\hline
$Model :$  & $1a$ & $1b$ & $2a$ & $2b$  & $4$ & $6$ \\
\hline \hline
$\Sigma_0(M_{\odot}/{pc^2})$ & 50 &100 &50 &100 &50 & 80 \\
$R_d (kpc) $ & 3.5 &3.5& 3.5 & 3.5  & 3.5  & 3.  \\
\hline
$R_c (kpc) $ & 5 & 5 & 5 & 5 & 5 & 15  \\
$\rho_{\odot}$ &0.008 & 0.008 & 0.008 & 0.003 & 0.007& 0.005 \\
$\beta$ & - & - & 0 & 0 & 0.2 & 0  \\
$q$ & - & - & 1 & 1 & 1 & 1 \\
$V_a$  & - & - & 165 & 100 & 170 & 170 \\
$M_{Halo}(60kpc)(10^{11}M_{\odot})$ & 5.1 & 5.1 & 1.9 & 0.7 & 1.2
& 2.2  \\ \hline \noalign{\smallskip}
\end{tabular}
\label{tmodel}
\end{center}
\caption{ Parameters of the power law model. First part of the
table represents the parameters of the disk and the second part
shows the halo parameters.}
\end{table*}
Unlike the Evans models, in the standard halo model $(1a,1b)$,
the velocity dispersion of lenses in the halo is approximately
independent from space and its value  $\sigma = 156 km/sec$.
Figure (\ref{vel1}) shows the transverse velocity distribution in
the spherically symmetric models 2a, 2b, 4 and 6.
\begin{figure}[htbp] 
\vspace*{8pt} \centerline{\psfig{file=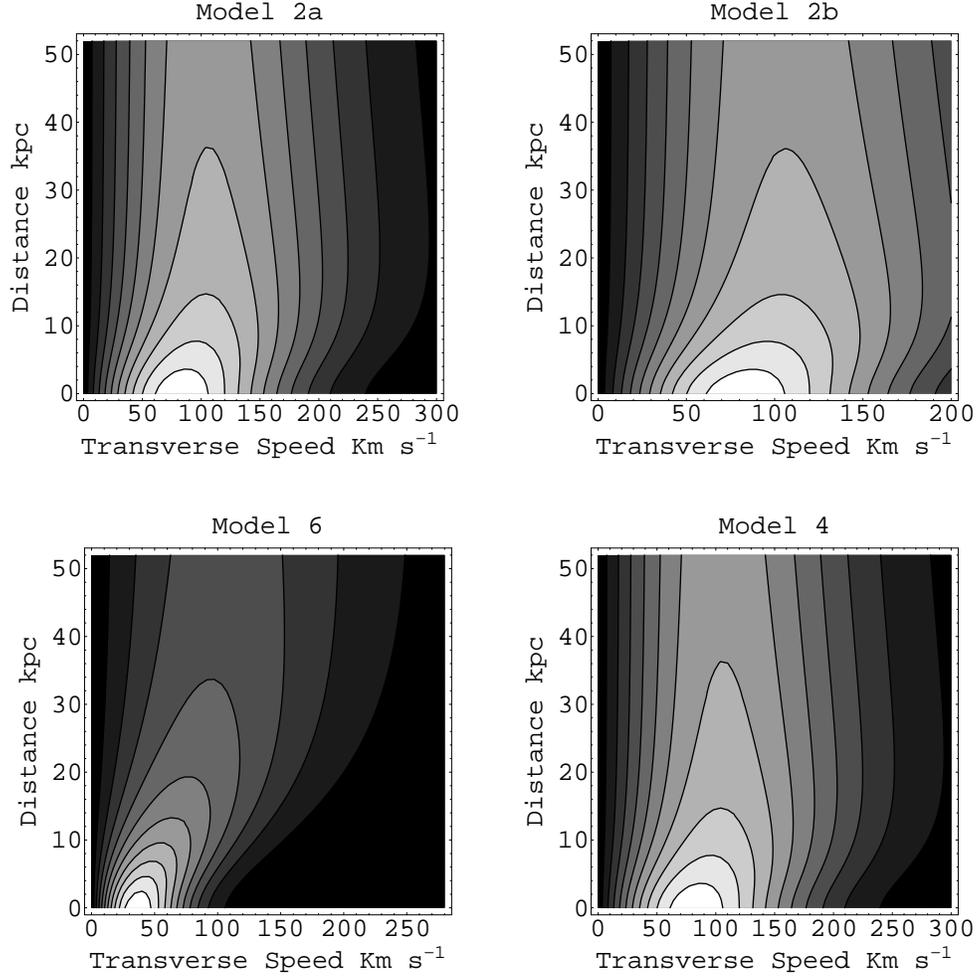}} \vspace*{8pt}
 \caption{2-Dimension Contours shows the distribution of transverse velocity and the
distance of lenses with respect to an observer that located at the
position of sun and observes the direction LMC in different Evans
halo models.} \label{vel1}
\end{figure}
Next section contain Monte-Carlo simulation for generating
microlensing events in Evans model to evaluate the expected rate
of events.
\section{ Monte-Carlo Simulation}
In this section we use Monte-Carlo simulation to obtain the
theoretical distribution of events duration for each galactic
models. We then apply the observational efficiency of EROS and
MACHO groups to the theoretical distributions to obtain the
expected distribution of events. One can compare these
distributions with the observed events and the aim is to put a
constraint on the fraction of halo that can be made by MACHOs.
Here, we supposed that all the microlensing events are due to
halo or galactic disk lenses and we ignore those located in the
LMC itself (self lensing). Next generation microlensing
experiments with high sampling rate and sufficient photometric
precision, definitely will solve the question of self-lensing
hypothesis \cite{rah02}.\\
We introduce the main functions for this Monte-Carlo simulation
which are the spatial distribution of lenses, distribution of
velocity and the mass function of compact objects. For the mass
function of the halo, we use the power law model as follows
\cite{shu96}:
\begin{equation}
P(M) d\left(\frac{M}{M_0}\right) =
A\left(\frac{M}{M_0}\right)^\alpha d\left(\frac{M}{M_0}\right),
\hspace{0.5cm} for \hspace{0.5cm} M_{min}\leq M \leq M_{max},
\label{massfunc}
\end{equation}
where $M_0= (M_{min}M_{max})^{1/2}$. The exponent $\alpha = -1.5$
according to the Expression (\ref{dn}) corresponds to an equal
rate of microlensing events per decade of lens masses and also
$\alpha = -2$ corresponds to an equal contribution to the optical
depth per decade of less masses. In the range of
$-1.5<\alpha<-2$, the optical depth is dominated by massive
objects and event rate is dominated by low mass objects. For the
$\alpha <-2$ both optical depth and event rate are dominated by
low massive objects, while for $\alpha>-1.5$ optical depth and
event rate are dominated by massive objects. The domain for the
Mass Function is defined by $\beta$:
\begin{equation}
\beta = \log (M_{max}/M_{min})
\end{equation}
Here in this simulation we choose $\alpha = -1.5$ and $\beta = 1$.
One can simplify problem by identifying a mass scale. We assume
such a mass scale is provided by a fixed upper mass limit, say
$M_{max} = 1 M_{\odot}$. In the case of mass function for the
disk, it has been proposed by HST observations \cite{gou97}. In
the disk MF the slope is changed at $M \sim 0.6 M_{\odot}$, from
a near-Salpeter power-law index of $\alpha = -1.21$ to $\alpha =
0.44$. The best fit to mass function of disk indicated straight
line , $d\log N/d\log M = -1.37 - 1.21 \log (\frac{M}{M_{\odot}})$
for $M>0.6M_{\odot}$ and $d\log N/d \log M = -0.99 +0.44
\log(\frac{M}{M_{\odot}})$ for $M<0.6M_{\odot}$.

\begin{figure}[htbp] 
\centerline{\psfig{file=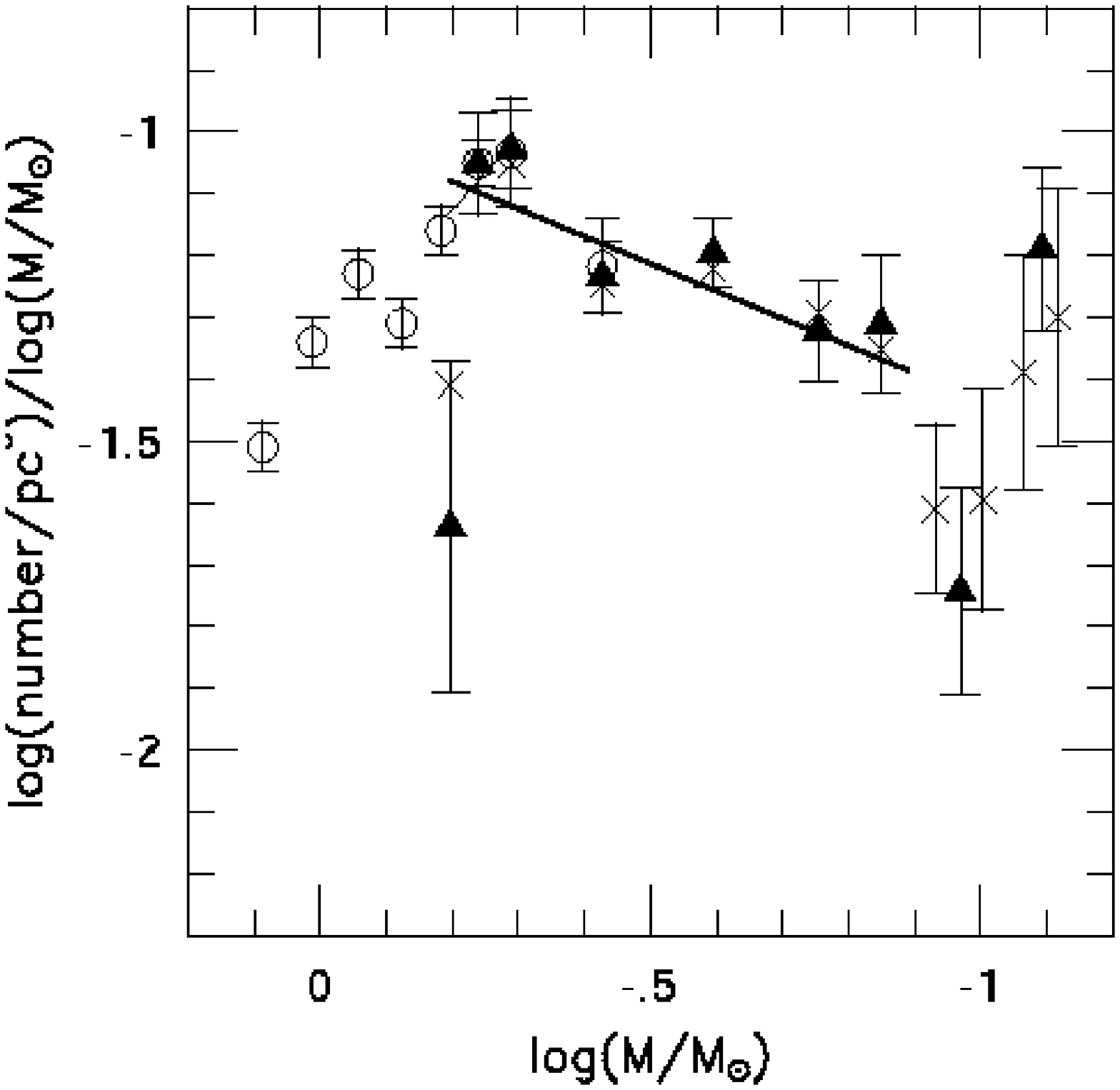,angle=0,width=12.cm,clip=}}
\caption{The mass function of disk \cite{gou97}. The slope of the
MF changes near the $M \sim 0.6 M_{\odot}$, from a near-Salpeter
power-law index of $\alpha = -1.21$ to $\alpha = 0.44$. }
\label{mfhst}
\end{figure}

Figure (\ref{mfhst}) shows the observed mass function by HST.\\
The spatial distribution of lenses obtain along our line of sight
by Eq. (\ref{dn})
$$probability \hspace{0.2cm} of \hspace{0.2cm} observation \propto \rho(x)\sqrt{x(1-x)}.$$
Using the different galactic models mentioned in Table.(1), the
spatial distribution of lenses along our line of sight obtain
according to Figures (\ref{rhohalo} \& \ref{rhodist}).
\begin{figure}[htbp] 
\vspace*{10pt} \centerline{\psfig{file=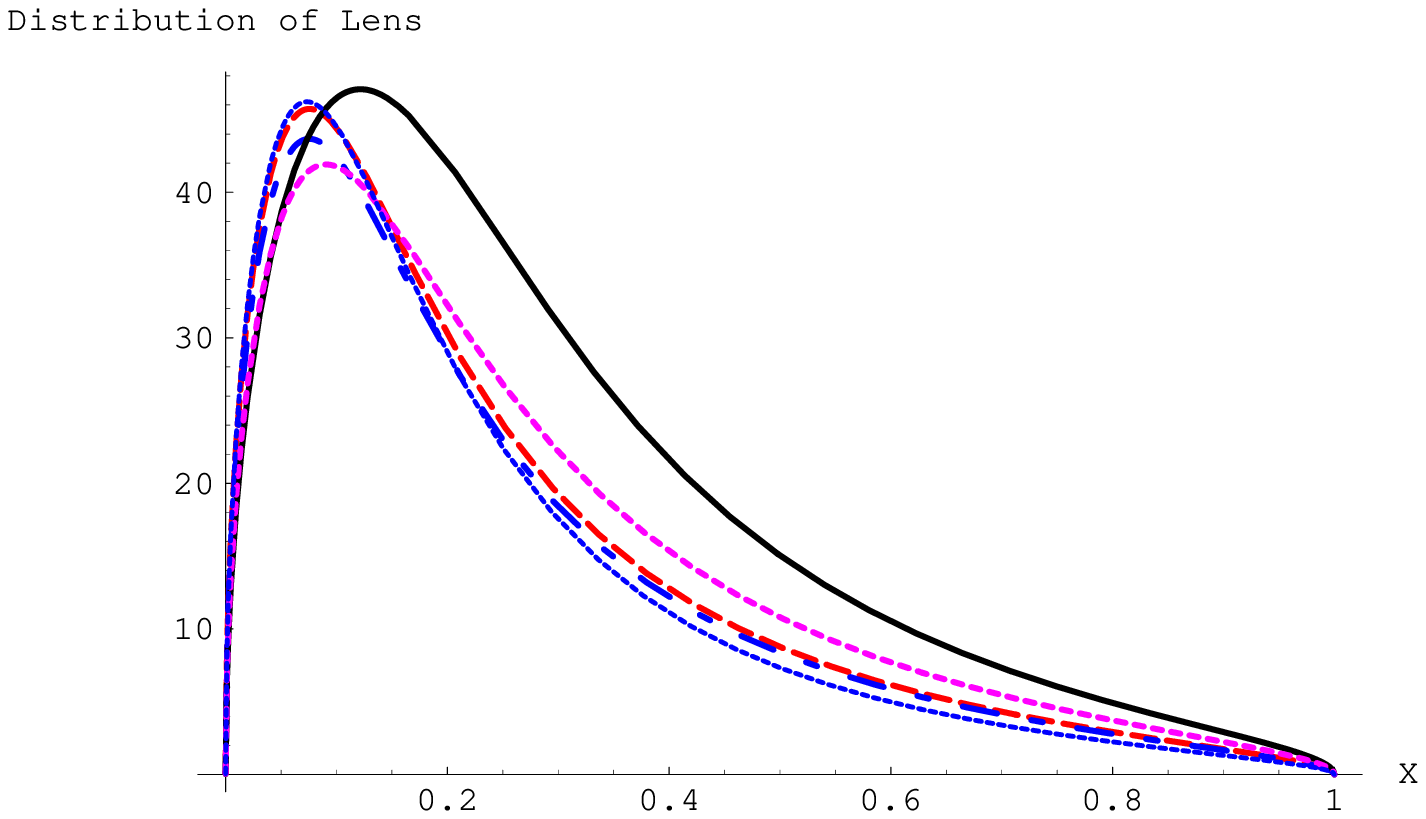}}
\vspace*{10pt}
\caption{The distribution of lenses as function of
distance from the observer in the different galactic halo models.
dot-line stands for standard halo, dot-thin line for model 4,
thick line for model 6, dot-dashed line for model 2b and dashed
line represents model 2a. } \label{rhohalo}
\end{figure}

\begin{figure}[htbp] 
\vspace*{10pt} \centerline{\psfig{file=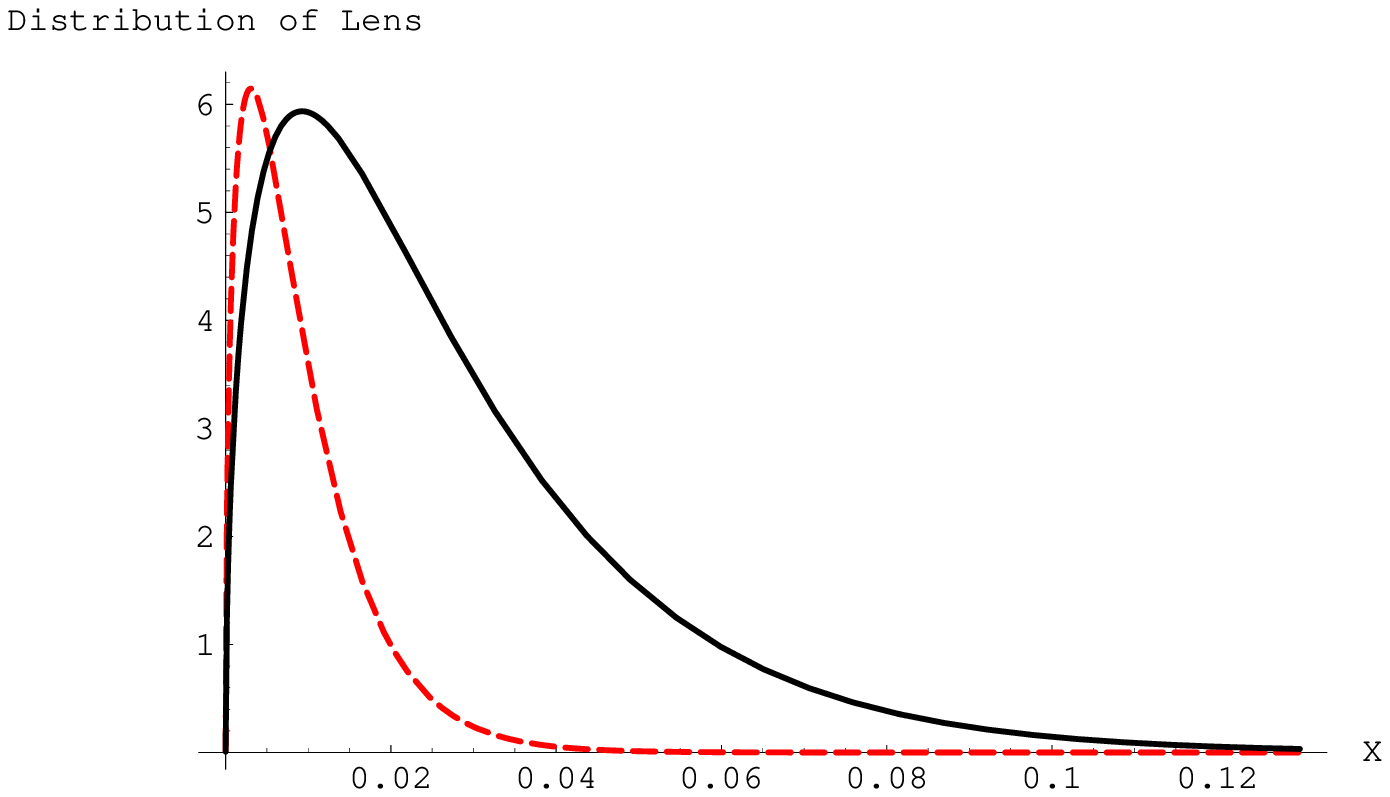}}
\vspace*{10pt}
 \caption{The distribution of lenses as function of distance from the observer in the different
 disk models. Solid-line represents this distribution in thick disk model
 and dashed line shows thin disk model.}
\label{rhodist}
\end{figure}
Here is the algorithm that we use in this simulation: \\
We start from the distance of the lenses from the observer via the
distance distributions, in Figures.( \ref{rhohalo} \&
\ref{rhodist}). The mass of MACHOs obtain according to the mass
functions of the Halo and disk. Combing the distance and the mass
of the lens by Eq. (\ref{einr}) yields the corresponding Einstein
radius of the lens. The transverse velocity distribution also
depending on the galactic models, has been obtained in Figures
(\ref{vel2} \& \ref{vel1}). We use the transverse velocity of the
lens and its Einstein radius to calculate the duration of events
by following formula:

\begin{equation}
\label{dur}
 t_E = 78.11
\left(\frac{M}{M_{\odot}}\right)^{1/2}\left(\frac{D_s}{10
kpc}x(1-x)\right)^{1/2}\left(\frac{ 200 km/s}{v_t}\right)
\hspace{0.5cm} days
\end{equation}

\begin{figure}[htbp] 
\centerline{\psfig{file=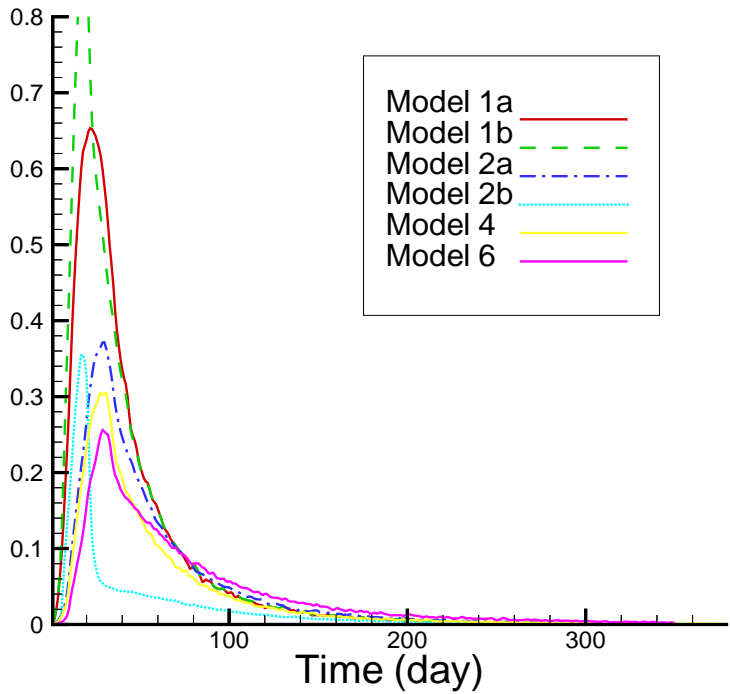,angle=0,width=12.cm,clip= }}
 \caption{ The distribution of events verses duration of events for
 different galactic models. The total number of events are normalized with the rate of
 events in one year for $10^7$ background stars which obtained by Table \ref{lmc}. }
\label{theomodel}
\end{figure}
Figure (\ref{theomodel}) shows the distribution of events for
different galactic models, where the number of events are
normalized $10^7 star \times yr$ exposure time with $100$ percent
efficiency in observation. The normalization have been done by Eq.
(\ref{gam4}), using the numerical optical depth value and the mean
of inverse Einstein crossing time.\footnote{We used THalo code
developed in EROS for optical depth calculation} Here we assume
that hundred percent of halo is made by compact halo objects.
Table (\ref{lmc}) as result of simulation shows corresponding
optical depth and the rate of events in galactic models.
\begin{table*} \caption[]{ Optical depth and the rate of
microlensing events is given toward LMC for each elements of
galactic structure.}
\begin{center}
\begin{tabular}{lclclclclclclc|}
\hline
$Model :$  & $1a$ & $1b$ & $2a$ & $2b$ & $4$ & $6$ \\
\hline \hline
$\tau_{Halo}(LMC)10^{-7}$ & 4.96   &  4.96 & 3.92 &1.44 &2.82 & 3.58\\
$\tau_{Disk}(LMC)10^{-7}$  &0.19  &0.39 & 0.19 & 0.39 & 0.19 & 0.31   \\
$\tau_{Total}(LMC)10^{-7}$ & 5.15 & 5.35 & 4.11 & 1.83 & 3.1 & 3.89 \\
\hline
$\Gamma_{Halo}(/10^7 star Yr)(LMC)$ &41.5& 41.5& 18.21& 5.53 & 16.38& 15. \\
$\Gamma_{Disk}(/10^7 star Yr)(LMC)$ &2.20  &4.40 &2.20 & 4.40 & 2.20 & 3.52    \\
\hline
$\Gamma_{Total}(/10^7 star Yr)(LMC)$ &43.69  & 45.89 &20.41 & 9.93 & 18.58 & 18.52  \\
\end{tabular}
\end{center}
\label{lmc}
\end{table*}
To obtain quantitative conclusion, we clearly need to assess our
event detection efficiency. The detection probability for
individual events is complication function of $u_0$, $t_E$ , the
strategy of observation and the brightness of the source stars. In
fact all these distributions except $t_E$ are not known and thus
can be averaged over by a Monte-Carlo simulation. Here we use the
observational efficiency of EROS and MACHO experiments in our
simulation. The detection efficiency of MACHO group in the term
of event duration can be found in \cite{suth99} and for the EROS
group in \cite{pal98}. For the efficiency calculation in EROS
group, microlensing parameters are drawn uniformly in the
following intervals: time of maximum magnification $t_0$ within
the observing period $\pm 150 days$, impact parameter normalized
to Einstein radius $u_0\in [0,2]$ and time scale $t_E\in[5,300]
days$ \cite{pal98}, where the efficiency is normalized to events
with the impact parameters $u_0<1$. The Efficiency of these two
groups are shown in Figure. (\ref{eff}). It should be mentioned
that the conventional definition of Einstein crossing time by
EROS is the half of its value defined by MACHO group, here by
convention we follow the time scale definition according to
equation (\ref{dur}).\\
\begin{figure}[htbp] 
\centerline{\psfig{file=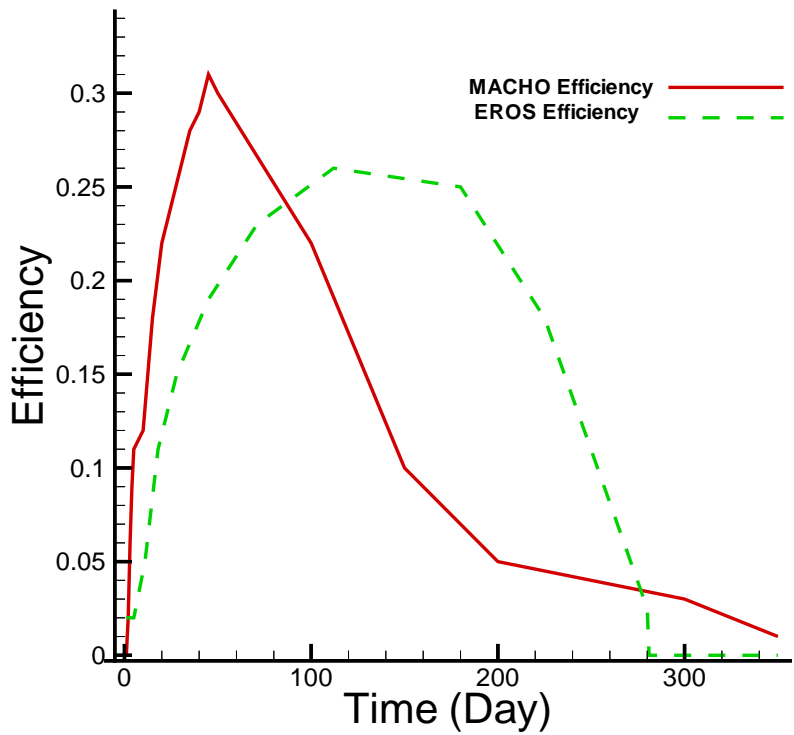,angle=0,width=12.cm,clip=}}
 \caption{ Detection efficiency verses event duration $t_E$ toward
 LMC. The solid line shows the efficiency of MACHO
 \cite{suth99} and dashed line indicates for EROS group \cite{pal98}.}
  \label{eff}
\end{figure}
The expected distribution of events can be obtained by
multiplying the efficiency to the theoretical distribution of
$d\Gamma/dt_E$. The expected observation is shown in Figure
(\ref{model}). Table (\ref{rate}) indicates overall expected
number of events by EROS and MACHO for $10^7$ time-object
exposure in each galactic models.
\begin{figure}
\centerline{\psfig{file=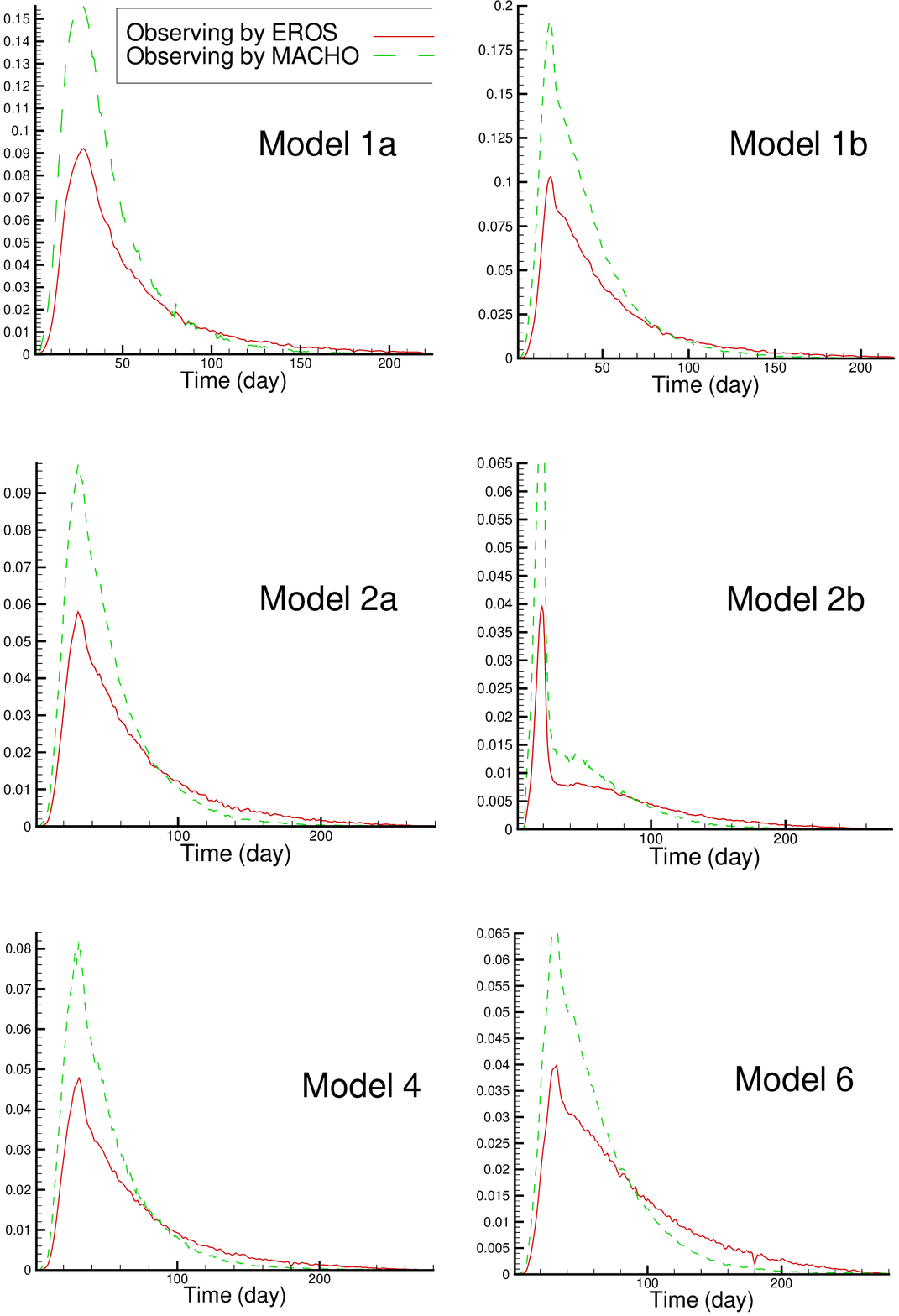,angle=0,width=12.cm,clip=}}
 \caption{
 The distributions of events verses duration are shown for
 different galactic models. The total number of events are normalized by Table (\ref{lmc}) with the
 exposure time of $10^7$ background stars in one year.}
\label{model}
\end{figure}
The Constraint on the contribution of the compact objects in the
dark matter of halo obtain by comparing the number of
microlensing candidates with those expected from galactic models.
Here we use two different statistical approach for analyzing EROS and MACHO data. \\
EROS observed two candidates during $ 2 year$ observations of
$17.5$ million stars in LMC \cite{las00}. The microlensing
candidates were called EROS-LMC-3 and 4 with Einstein crossing
time in $41$ and $106$ days. Now, for statistical analysis, let
${\it b}$ is the total expected number of events for a given
galactic model (Table \ref{rate}) and $n_0$ be the observed events
in the absence of background. Considering the Poisson probability
distribution for events, one can obtain Poisson confidence
intervals $[\mu_1,\mu_2]$ for $n_0$ observed events. In
particular case we want to put an upper limit for $b$ with a
certain level of confidence. For instance, observing two events
by EROS allows us to put a constraint, excluding at $95\%$ C.L
that the expected number of events should not be more than $6.72$
, let call this number "$C$".
To compare this number with Table. \ref{rate} we normalize the
numbers of table to the exposure time of EROS, $(2\times17.5\times
10^6 year*star)$. The fraction of halo that made by MACHOs can be
obtained by $f_M < C/F_M$ (at $95 \% CL)$ , where the $F_M$ is
the expected number of events in the Table. (\ref{rate}), taking
into account the normalization factor. The results for EROS
indicated in Table (\ref{frac}). It is seen that except the
minimal halo model, the compact objects form a fraction of halo mass. \\
The analysis of $5.7$ year of photometric on $11.9$ million stars
by MACHO group in the LMC also reveals 13-17 microlensing events
\cite{alc00}. We normalize the rate of events in the Table
(\ref{rate}) to the exposure time of MACHO, $(5.7 \times
11.9\times 10^6 star\times year)$. For the statistical analyzing
of MACHO results, unlike the former approach used for EROS
results, here we consider the part of halo mass that is allowed
to be in the form of MACHOs with one sigma error. This fraction
obtain by dividing $(13-17)$ events to the normalized number of
events in Table. (\ref{rate}). The results indicated in Table.
(\ref{frac}).

\begin{table*}
\caption[]{EROS/MACHO expected number of microlensing events,
considering that $100\%$ of halo made by compact objects are
obtained in $10^7 star\times yr$ exposure time.}

\begin{center}
\begin{tabular}{lclclclclclclc|}
\hline \noalign{\smallskip}
$Model :$  & $1a$ & $1b$ & $2a$ & $2b$ & $4$ & $6$ \\
\hline
{EROS}  & 6.81 & 6.93 & 3.64 & 1.59 & 3.29 & 3.23  \\
{MACHO} & 10.1 & 10.43 & 4.84 & 2.02 & 4.39 & 3.82  \\
\noalign{\smallskip} \hline
\end{tabular}
\end{center}
\label{rate}
\end{table*}

\begin{table*}
\caption[]{The fraction of halo in the form of MACHO obtain by
comparing the observational results of experiments with the
expected theoretical number of events. This fraction depends on
galactic model and the mass function of compact halo objects. The
EROS results obtain by excluding $f_{EROS}$ with $ 95 \% C.L$ to
be less than indicated value. MACHO results have also been
obtained with one sigma error for $f_{MACHO}$.}
\begin{center}
\begin{tabular}{lclclclclclclc|}
\hline
\noalign{\smallskip}
$Model :$  & $1a$ & $1b$ & $2a$ & $2b$ & $4$ & $6$ \\
\hline
{\bf $f_{EROS}<$}  & 0.28 & 0.27 & 0.52 & 1. & 0.58 & 0.59  \\
{\bf $f_{MACHO}$} & $0.25^{+0.06}_{-0.06}$  & $0.24^{+0.05}_{-0.05}$ & $0.52^{+0.12}_{-0.12}$ & 1 & $0.57^{+0.14}_{-0.14}$ & $0.66^{+0.16}_{-0.16}$  \\
\noalign{\smallskip} \hline
\end{tabular}
\end{center}
\label{frac}
\end{table*}
\section{Conclusion}
Two years observation of Large Magellanic Clouds by EROS and 5.7
years by MACHO, revealed $2$ and $(13-17)$ microlensing
candidates, respectively. The results presented here provide some
interesting conclusion on the contribution of MACHOs in the halo
of our galaxy, depending on the models.\\
It is seen that except minimal halo model, the number of observed
events are inadequate that halo fully comprised of $[0.1, 1]
M_{\odot}$ compact halo objects. Two extreme results are
considering a non spherical halo (minimal halo) and another
possibility is an LMC halo that dominate microlensing, and no
MACHOs in the Milky Way. The essential way to distinguish the real
model of our galaxy and the contribution of compact objects on
the halo is localizing the position of lenses. Studying parallax,
finite size effect and double lenses would allow us to achieve
this aim. Recently, analysis of MACHO-99-BLG-22/OGLE-1999-BUL-32,
indicated parallax effect in its light curve. A likelihood
analysis of the lens position implies that lens could be a black
hole \cite{ben02} \& \cite{mao02}. The analysis of PLANET
photometric observations of event EROS BLG-2000-5
\footnote{http://www-dapnia.cea.fr/Spp/Experiences/EROS/aletrs.html}
shows the parallax and binary orbital motion in its light curve.
It is the first time that the lens mass degeneracy have been
completely broken \cite{an02}. In spite of localizing some events
toward galactic bulge, in the direction of LMC, the position of
lenses have not been localized and it needs next generation
microlensing experiments with high sampling rate and better
photometric
precision \cite{rah02}.\\
The fraction of halo (derived from microlensing experiments) in
the form of MACHOs can also be compared with the White Dwarf
\cite{iba99} \& \cite{men99} population from the Hubble Deep
Field. Although the identification of these faint blue objects as
white dwarfs remains to be confirmed and the small sample size
restricts an accurate estimate, the suggestion that these White
dwarfs could contribute $1/3$ to $1/2$ of the dark matter in the
Milky Way. This result is in agreement with the fraction of halo
in the form of compact objects in some of halo models. The
galactic halo, composed of White Dwarfs would seem to be a natural
explanation of the microlensing data.





\begin{thebibliography}{99}
\bibitem{fab79}
Faber, S. M., Gallagher J. S., 1979,  Ann. Rev. Astron. Astrophys
17, 135.

\bibitem{tri87}
Trimble, V., 1987, Ann. Rev. Astron. Astrophys 25, 425.

\bibitem{per96}
Persic, M., Salucci, P and Stel, F., 1996, MNRAS 281, 27.

\bibitem{moo94}
Moore, B., 1994, Nature 370, 629.

\bibitem{nav96}
Navarro, J. F., Frenk, C. S., White, S. D., 1996, APJ 462, 563.

\bibitem{cop95}
Copi, C. J., Schramm, D. N and Turner, M. S., 1995, Science 267,
192.

\bibitem{bur98}
Burles, S., Tyler, D., 1998, APJ 499, 699.


\bibitem{fuk95}
Fukugita M., Hogan C. J., Peebles P. J. E., 1995, A\&A 503, 518.

\bibitem{pac86}
Paczy\'nski B., 1986, APJ 304, 1.

\bibitem{eva94}
Evans N. W., 19914, MNRAS 267, 333.

\bibitem{spi01}
Spiro M., Lasserre T, {\it Cosmology and Particle Physics},
edited by Durrer, R., Garcia-Bellido, J and Shaposhnikov, M
(American Intitute of Physics),(2001) 146.


\bibitem{alc00}
Alcock C. et al. (MACHO)., 2000, APJ 542, 281.

\bibitem{pac96}
Paczy\'nski B., 1996, Annu. Rev. Astron. Astrophys 34, 419.


\bibitem{rou97}
Roulet E., Mollerach S., 1997, Phys. Rep 279, 67.

\bibitem{gou00}
Gould A, {\it A New Era of Microlensing Astrophysics},edited by
Menzies J. W., Sackett P. D., (ASP conference Series), preprint
(astro-ph/0004042).

\bibitem{jet97}
Jetzer P., 1997, Proceeding of {\it 8th Marcel Grossmann Meeting
on Relativistic Astrophysics}, Jerusalem, preprint
(astro-ph/9709212).


\bibitem{bin87}
Binney S., Tremaine S {\it Galactic Dynamics}. Princeton
University Press (1987).


\bibitem{alc94}
Alcock C. et al. (MACHO), 1995, APJ 449, 28.

\bibitem{ren96}
Renault C., PhD thesis, Universit\'e, Paris 7, DAPNIA/SPP
96-1003, (pub no. 96001264)

\bibitem{pal97}
Palanque-Delabrouille, PhD thesis, University Paris 7 and
University of Chicago, DAPNIA/SPP 97-1007

\bibitem{rah02}
Rahvar S., Moniez M., Ansari R., Perdereau O., 2002 (submitted in
A\&A).


\bibitem{shu96}
Mao S., Paczynski B., 1996, APJ 473, 57.

\bibitem{suth99}
Sutherland W., 1999, Rev. Mod. Phys. 71, 421.

\bibitem{pal98}

Palanque-Delabrouille N. et al. (EROS)., 1998, A\&A 332, 1.

\bibitem{gou97}
Gould A., Bahcall J. N., Flynn C., 1997, APJ 482, 913.

\bibitem{las00}
Lasserre T. et al. (EROS)., 2000, A\&A 355, 39.

\bibitem{ben02}
Bennett, D. P. {\it et al}, preprint (astro-ph/0207006).

\bibitem{mao02}
Mao, S. {\it et al}., 2002, MNRAS, 329, 349.

\bibitem{an02}
An, J. {\it et al}., 2002, APJ, in press (astro-ph/0110095).

\bibitem{iba99}
Ibata, R. A. {\it et al}., 1999, APJ 524, 1.

\bibitem{men99}
Me\'ndez, R. A and Minniti, D., 2000, APJ 529, 911.

\end {thebibliography}
\end{document}